\documentstyle[aps,12pt]{revtex}

\begin{document}
\title{Loss of the magnetism in FeS }
\author{Z. Ropka}
\address{Center for Solid State Physics, S$^{nt}$ Filip 5, 31-150 Krak\'{o}w.%
}
\author{R.J. Radwanski}
\address{Center for Solid State Physics, S$^{nt}$ Filip 5, 31-150 Krak\'{o}%
w, \\
Institute of Physics, Pedagogical University, 30-084 Krak\'{o}w, POLAND.\\
email: sfradwan@cyf-kr.edu.pl}
\maketitle

\begin{abstract}
It is argued that the formation of the non-magnetic state of FeS in external
pressures above 6.5 GPa is in agreement with the prediction of the quantum
atomistic solid-state theory, an extension of the crystal-field theory with
the spin-orbit coupling incorporated, and is related with the change of sign
of the off-cubic trigonal distortion marked by the change of the c/a ratio
from 
\mbox{$>$}%
1.63 (magnetic) to 
\mbox{$<$}%
1.63 (non-magnetic).



Receipt date by Phys.Rev.Lett. 29.08.2000
\end{abstract}

\date{(29.08.2000)}

Recently pressure-induced high-spin to low-spin transition has been
evidenced in FeS [1]. A pressure of 6.5 GPa destroys the quite strong
magnetism of FeS. FeS is\ an antiferromagnet with the relatively high Neel
temperature of 600 K. It has the hexagonal NiAs structure with a=345 pm and
c=585 pm at room temperature [2].

The aim of this Letter is to point out that the loss of magnetism of FeS can
be understood within the quantum atomistic solid-state theory (QUASST), an
extension of the crystal-field theory with the spin-orbit coupling
incorporated, that points out the existence of the discrete atomic-like
states in 3d-ion containing compounds associated with the atomic-like states
of the highly-correlated 3d$^{n}$ electron system. Within this theory it has
been proven that it is the local symmetry that determines the realization of
the low- or high-spin state of the Co$^{3+}$ ion [3-5]. Namely, it has been
proven that as the function of the sign of the off-cubic trigonal distortion
of the octahedral crystal-field the ground state of the Co$^{3+}$ ion is the
magnetic doublet (the elongation along the main diagonal of the octahedron)
or the non-magnetic singlet (the stretching).

Exactly such the conditions are realized in FeS. We learned about it when
Ref. 2 came to our attention.

The results obtained in Refs 3 and 4 for the Co$^{3+}$ ion are directly
applicable to the Fe$^{2+}$ ion as both systems have 6 electrons in the open
3d-shell in the form of the highly-correlated 3d$^{6}$ system. Analyzing the
hexagonal NiAs-type structure of FeS one finds that the Fe$^{2+}$ cation in
FeS is in the quasi-octahedral site of the S$^{2-}$ anions. For the ideal
value c/a of $\frac{2}{3}\sqrt{6\text{ }}$ = 1.63 the surrounding is the
exact octahedron. The beauty of the hexagonal NiAs structure relies in a
fact that the trigonal off-cubic distortion can be realized without the
spoiling of the overall hexagonal symmetry and the distortion direction is
along the hexagonal c axis. The macroscopic lattice parameters reflect
directly the atomic-scale octahedron. The c/a 
\mbox{$>$}
1.63 corresponds to the trigonal elongation of the atomic-scale octahedron
whereas the trigonal stretching is marked by c/a 
\mbox{$<$}%
1.63.

According to Ref. 2 with the increasing pressure the ratio c/a stays almost
constant ($\sim $1.70) up to 6.5 GPa. At 6.5 GPa it undergoes the abrupt
decrease of the lattice parameter c, by 8.5\%, and c/a becomes 1.58 only.
Hence, during the transition the sign of the off-cubic trigonal distortion
changes. The distortion crystal-field parameter B$_{2}^{0}$, discussed in
Refs 3-5, is in the first-order approximation proportional to B$_{2}^{0}\sim 
$(1.633-c/a) (for the 3d$^{6}$ system).

It is worth to add that the non-magnetic state of FeS can be regarded as the
state when the Jahn-Teller effect wins the magnetic state. In fact, the
non-magnetic state of the Fe$^{2+}$ ion in FeS can be regarded as quite
normal owing to the fact that the Fe$^{2+}$ ion is the non-Kramers system (3d%
$^{6}$). According to the Jahn-Teller theorem [6] and Kramers theorem the
ground state should be a non-magnetic singlet produced by lattice
distortions. Thus, it is magnetic Fe$^{2+}$-ion compounds that should be
treated as exceptional. The quite common magnetism of the Fe compounds is
related with the existence of a great number of closely lying localized
states at the neighborhood of the ground state. Such the situation helps
development of magnetism as we know from rare-earth compounds. The
significant number of closely-lying localized states is somehow related to
the small value of the intra-atomic spin-orbit coupling in case of 3d ions
[7].

In conclusion, we argue that the formation of the non-magnetic state of FeS
in external pressures above 6.5 GPa is in agreement with the prediction of
the quantum atomistic solid-state theory (an extension of the crystal-field
theory with the spin-orbit coupling incorporated) and is related with the
change of sign of the off-cubic trigonal distortion marked by the change of
the c/a ratio from 
\mbox{$>$}%
1.63 (magnetic) to 
\mbox{$<$}%
1.63 (non-magnetic). The fine electronic structure of FeS in the magnetic
state (low pressure) is quite similar to that derived for FeBr$_{2}$ [8]
with the magnetic doublet ground state. At high pressures the arrangement of
the doublet-singlet sequence reverses to the fine electronic structure found
in LaCoO$_{3}$ [3,5] with the non-magnetic singlet ground state.

E-mail for correspondence: sfradwan@cyf-kr.edu.pl

\end{document}